\documentclass{emulateapj}
\usepackage{apjfonts}

\submitted{Received 2007 July 11; accepted 2007 August 1}
\journalinfo{The Astrophysical Journal, 667:L?--L?, 2007 September 20}

\def\mV{$m_{\rm F606W}$}
\def\mI{$m_{\rm F814W}$}
\def\hi{H{\sc i}}

\begin{document}


\title{Stellar populations across the NGC\,4244 truncated galactic disk}


\author{Roelof S.\ de Jong\altaffilmark{1}, A.C.\
  Seth\altaffilmark{2}, D.J.\ Radburn-Smith\altaffilmark{1}, E.F.\
  Bell\altaffilmark{3}, T.M.\ Brown\altaffilmark{1}, J.S.\
  Bullock\altaffilmark{4}, S.\ Courteau\altaffilmark{5}, J.J.\
  Dalcanton\altaffilmark{6}, H.C.\ Ferguson\altaffilmark{1}, P.\
  Goudfrooij\altaffilmark{1}, S.\ Holfeltz\altaffilmark{1}, 
  B.W.\ Holwerda\altaffilmark{1}, C.\
  Purcell\altaffilmark{4}, J.\ Sick\altaffilmark{7}, D.B.\ Zucker\altaffilmark{8}}

\altaffiltext{1}{Space Telescope Science Institute, Baltimore, MD 21218}
\altaffiltext{2}{Harvard-Smithsonian Center for Astrophysics, Cambridge, MA 02138}
\altaffiltext{3}{Max-Planck-Institut f\"ur Astronomie, D-69117, Heidelberg, Germany}
\altaffiltext{4}{University of California, Irvine, CA 92697}
\altaffiltext{5}{Queen's University, Kingston, Ontario, Canada}
\altaffiltext{6}{University of Washington, Seattle, WA 98195}
\altaffiltext{7}{Rice University, Houston, TX 77005}
\altaffiltext{8}{University of Cambridge, Cambridge CB3 0HA, UK}

\begin{abstract}
  We use HST/ACS to study the resolved stellar populations of the
  nearby, nearly edge-on galaxy NGC\,4244 across its outer disk
  surface density break. The stellar photometry allows us to study the
  distribution of different stellar populations and reach very low
  equivalent surface brightnesses.
  We find that the break occurs at the same radius for young,
  intermediate age, and old stars. The
  stellar density beyond the break drops sharply by a factor of at least
  600 in 5 kpc.  The break occurs at the same radius independent of
  height above the disk, but is sharpest in the midplane and nearly
  disappears at large heights. 
  %This would reduce the sharpness of the
  %break if seen face-on.
%
  These results make it unlikely that truncations are caused
  by a star formation threshold alone: the threshold would have to keep
  the same radial position from less than 100 Myr to 10 Gyr ago, in
  spite of potential disturbances such as infall and redistribution of
  gas by internal processes. A dynamical interpretation of truncation
  formation is more likely such as due to angular momentum
  redistribution by bars or density waves, or heating and stripping of
  stars caused by the bombardment of dark matter sub-halos. The latter
  explanation is also in quantitative agreement with the small diffuse
  component we see around the galaxy.

\end{abstract}



\section{Introduction}

Using photographic plates of edge-on galaxies, \citet{vdK79} was the
first to note that some spiral galaxies have truncated disks. More
recent studies with CCDs of galaxies at a range of inclinations have
found that these truncations are not the end of the stellar disk, but
are better described as a change in the exponential length scale of
the radial light distribution \citep{Poh02}.  These downward breaks in
the light profile occur in the majority of galaxies \citep{PohTru06}.
Many different models have been proposed to explain the truncations or
breaks but they can all be placed into two broad categories: models
that form stars in the observed configuration and models that
transport stars into the observed configuration by dynamical
effects. Here we use observations of the resolved stellar populations
of the edge-on galaxy NGC\,4244 to constrain these types of models.

The observed projection of a galaxy can affect the detectability
of breaks in the light profiles.  Breaks were originally discovered
using edge-on galaxies \citep{vdK79,vdKSea81a} and seem to be more
readily observed in edge-on galaxies than face-on galaxies. This can
be partly explained by the line-of-sight projection that allows one to
probe fainter (de-projected) surface brightnesses for edge-on
galaxies. Furthermore, to reach faint isophotes in face-on galaxies,
azimuthal smoothing is used to extract radial profiles. This
technique, however, smooths out the non-axisymmetric features (bars, spiral
arms) that could result in sharp features in edge-on projections.

A number of recent studies with larger samples of galaxies have tried
to bridge the projection gap and have identified a number of
characteristics of breaks. \citet{Erw05} and \citet{PohTru06} identify
three types of luminosity profiles: those with a decreasing scale
length beyond the break, those with an increasing scale length, and
those with no break at all. Downward breaks generally occur at 2--4
disk scale lengths, with galaxies at the short end of that range
typically exhibiting lower central surface brightnesses and larger
scale lengths. These correllations cause the breaks to occur at a similar
surface brightness in different galaxies \citep{KrevdK04}. The breaks
occur at about the same radius independent of the height above the
midplane \citep{Poh07}, and appear to be present already
at redshifts z$\sim$1 \citep{Per04,TruPoh05}. There are some
indications that disk breaks and (gaseous) warps are related, but
there are exceptions to the rule \citep{vdK07}.

As mentioned earlier, models to explain the downward breaks in galaxy
disks come in two broad categories. In the first group the breaks are
created by stars forming intrinsically in the observed distribution,
either as a result of a limit in the gas distribution \citep{vdK87},
or some kind of threshold suppressing star formation
\citep[e.g.,][]{FalEfs80,Ken89, Sch04,ElmHun06}. In the second
category are models that rearrange the stars after they form. This may
occur by redistribution of angular momentum due to bars and spiral
arms \citep[e.g.,][Foyle et al., in prep.]{Deb06}, or by stripping of
stars due to repeated tidal interaction with other galaxies
\citep[e.g.,][]{Gne03} or dark matter sub-halos (Kazantzidis et al.,
in prep.).

To constrain the large number of models in the literature we
investigate the resolved stellar population along the major axis of
NGC\,4244 to study the age distribution of stars across its break.
NGC\,4244 is a late-type SAcd spiral galaxy with a maximum rotation speed of
about 95 km s$^{-1}$.  We use a distance modulus of 28.20 for NGC\,4244 as
determined from the tip of the Red Giant Branch (RGB) method by
\citet{Seth05I}.  
The observations and data reduction are described in
\S \ref{obsred}, the stellar distributions are investigated in
\S \ref{profs}, and we finish with discussion and conclusions in
\S \ref{concl}.

\section{Observations and Data Reduction}
\label{obsred}

Our stellar photometry was derived from HST observations in the F606W
and the F814W bands with the Advanced Camera for Surveys (ACS) Wide
Field Camera. We obtained SNAP observations
%observations for NGC\,4244 were SNAP observations in
%programs 9765 (central field) and 10523 (7 other fields) 
with typical total exposure times of about 700 s in each filter split
across two exposures. This data set is part of the GHOSTS
%\footnote{GHOSTS: Galaxy
%Halos, Outer disks, Substructure, Thick disks and Star clusters}
survey of 14 galaxies with similar data as presented here \citep{deJ07}.  
%The HST archive pipeline reduced flatfielded images were conservatively
%cleaned from cosmic rays using LA\_COSMIC (van Dokkum ***) and
%mosaiced together with MULTIDRIZZLE (Koekemoer ***). We used
%SExtractor (Bertin ***) on this mosaiced image to identify clearly
%extended objects and other image defects to create a mask that was
%used to exclude these areas from all photometry.

We used DOLPHOT \citep{Dol00} to obtain PSF-fitting stellar photometry
of our fields. 
%DOLPHOT uses the individual flatfield frames and not
%the stacked multidrizzle image. 
The output of DOLPHOT was filtered to select only real stars using a
combination of criteria on signal-to-noise (S/N), magnitude, and
DOLPHOT flag and sharpness parameters. These criteria remove most
non-stellar objects (galaxies and image defects) while leaving a large
fraction (but certainly not all) stellar objects in the sample.
Contamination from unresolved galaxies occurs only at low S/N, where a
sharpness criterion is too poorly defined to separate all galaxies
from stars. At the bright end there may be some Galactic foreground
contamination, but this is expected to be less than 10 stars per
3\arcmin$\times$3\arcmin\ ACS field for $m_{\rm F814W}$=23--26 mag
\citep{Gir05}.

We ran DOLPHOT artificial star tests with at least 1 million stars per ACS pointing to
calculate incompleteness corrections
as a function of stellar magnitude, color, and crowding. 
%On each of the ACS fields we inserted one million
%artificial stars spaced on a randomized grid and with colors and
%magnitudes 0.5 magnitude fainter than observed.
% using a color-magnitude distribution probability
%#distribution that mimicked the a smoothed observed distribution with
%an extension to fainter magnitude and with a floor probability over
%the full region of interest in the color-magnitude diagram. 
Each observed star was given an observing probability based on these
artificial star tests, which was used to correct the star count radial
profiles described in the next section. The average correction is
about 2.5 in the most crowded regions, reducing to less than 1.1 in
the outer regions. Full details of the GHOSTS data pipeline will be
described in a forthcoming paper (Radburn-Smith et al., in prep.).

\section{Stellar Density Profiles}
\label{profs}

\begin{figure}
\vspace{-1mm}
\includegraphics[width=\linewidth,trim=0mm 0mm 0mm 15mm,clip=true]{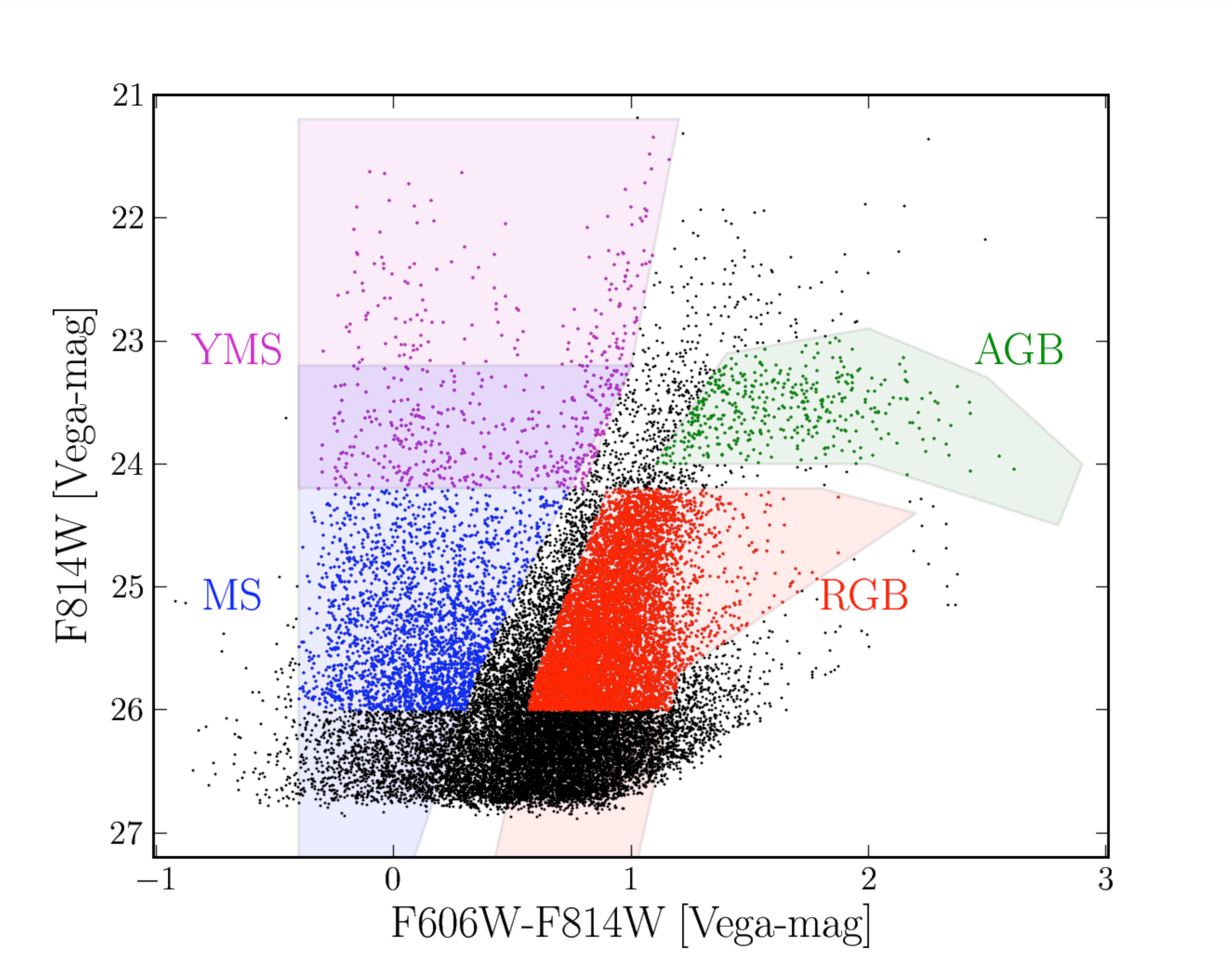}
\caption{Color-magnitude diagram of the ACS field straddling the
  North-East break of NGC\,4244. The colored polygons identify the
  marked populations. Colored points show the selected stars, limited
  to $m_{\rm F814W}<25.9$ mag.
\label{CMD}
}
\end{figure}

In Fig.~\ref{CMD} we show the color-magnitude diagram
for the ACS field at the North-East end of the disk that covers the disk
break. Several different stellar evolutionary stages can be identified
in this diagram. We have used polygon shapes to identify our
separation in broad evolutionary stages, but one should realize that
some overlap in age will exist in the different polygons for certain
ages and metallicities \citep[see also][]{Seth05II}.  
%Note that the ages 
%of each population box discussed below assume a constant star formation rate. 

The magenta polygon indicates the region dominated by stars less than
100 Myr old. The main sequence stars are on the left. Barely separated
from the main sequence at about \mV--\mI$\sim$0.1 are the blue He
burning stars. The spur up at about \mV--\mI$\sim$1.0 are red He
burnings stars. These stars have been partly excluded from this bin as they
overlap with the other populations. We refer to this bin as the YMS (young
main sequence) population.  The blue polygon is dominated by older
main sequence and He burning stars of about 100--300 Myr old (labeled
the MS population).
%At the faint end this polygon may have
%contamination of RGB stars and background galaxies. 
The green polygon indicates the region dominated by Asymptotic Giant
Branch (AGB) stars of at least 0.3 Gyr, but mostly of intermediate age
(1--3 Gyr), with a tail to older ages of up to 10 Gyr. The red polygon
is mostly populated by metal-poor RGB stars. Almost all these stars will
be older than 5 Gyr.  However, even for a constant star
formation rate population this polygon is dominated by $\sim$8--12 Gyr
old stars and younger populations can only dominate if star formation
was skewed toward younger ages.
%However, this region has contributions from younger He burning
%and AGB stars. 
We will use these four population boxes to investigate the
spatial distribution of stars of different age.

\begin{figure}
\includegraphics[width=\linewidth,trim=0mm 3mm 0mm 0mm,clip=true]{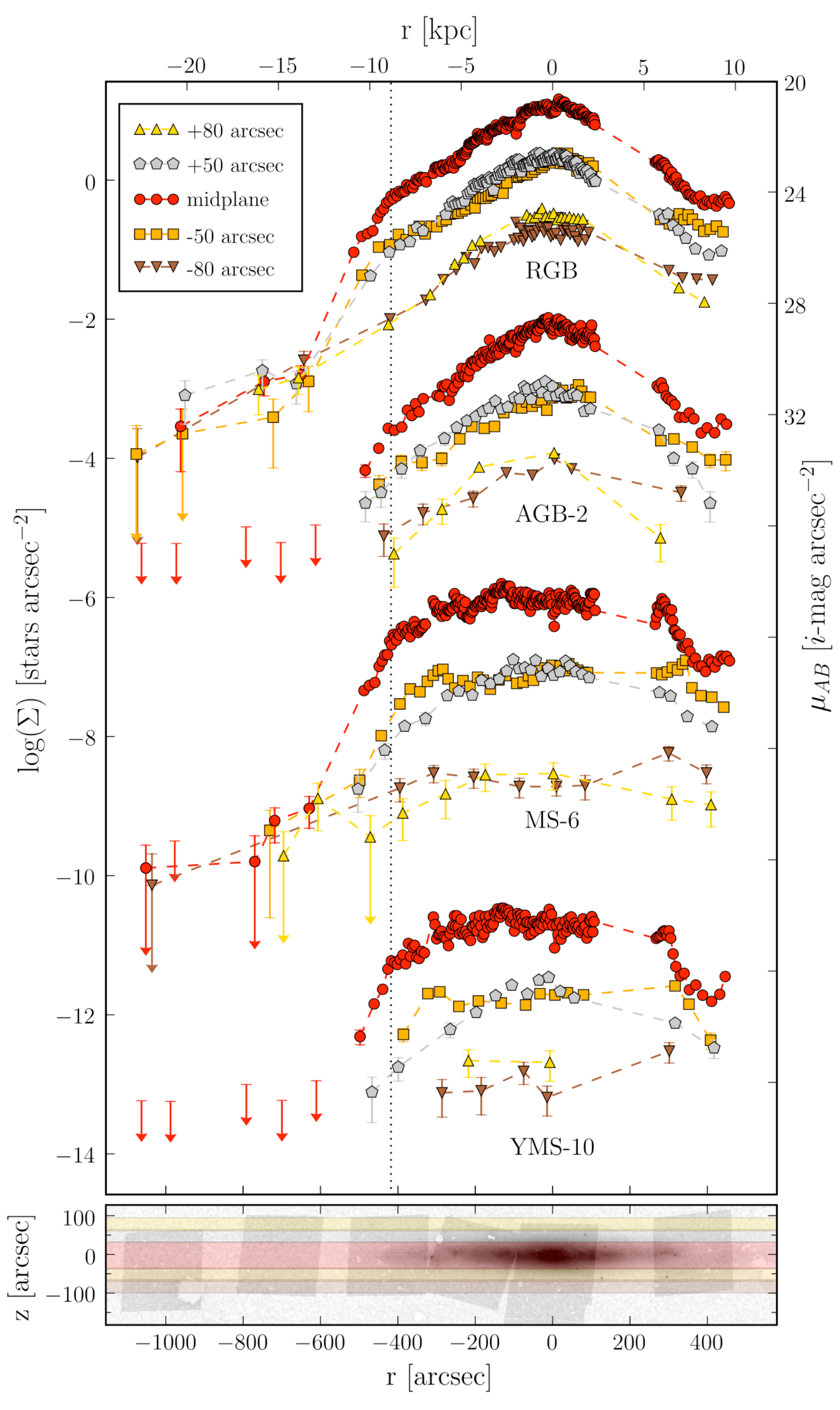}\\
\caption{{\it Top)} Incompleteness corrected star count profiles of
  NGC\,4244. Profiles are shown for a 70\arcsec\ wide strip along the
  midplane, and for 35\arcsec-wide strips offset 50\arcsec\ and
  80\arcsec\ above and below the plane. The different symbols and
  colors used for the 5 strips are identified in the
  legend. Contamination-subtracted star count surface density
  ($\Sigma$) profiles are presented for the four CMD regions
  identified in Fig.~\ref{CMD} as marked, where for clarity the AGB,
  MS, and YMS populations have been offset by $-2$, $-6$, and $-10$
  respectively in log($\Sigma$). The error bars are 2-sigma Poisson
  noise uncertainties, with arrows indicating that Poisson uncertainty
  will move the counts below the background level. Two-sigma upper
  limits are plotted for bins in which we detect fewer than 3
  stars. For reference we show the equivalent SDSS $i$-band surface
  brightnesses on the right axis, which was obtained by lining up the
  SDSS $i$-band integrated light profiles with the RGB stellar density
  profiles.  {\it Bottom)} SDSS image of NGC\,4244 rotated by
  42.7\degr, such that the North-East is to the left. Overlaid are the
  positions of our ACS fields in gray and the position of the strips
  used to extract the profiles above, in the same color coding.
\label{NGC4244majpop}
}
\end{figure}

In the bottom part of Fig.~\ref{NGC4244majpop} we show an SDSS image of
NGC\,4244 with the ACS fields overlaid. %The areas used to extract the
%star count and luminosity profiles are also shown.
% using the same color coding as used in the profile plots.
Shown in color are five strips used to extract star count profiles.
In the top part of Fig.~\ref{NGC4244majpop} we present the radial star
count in these five strips parallel to the major axis for our four
population boxes. The contamination subtracted star counts
are presented as symbols with 2-sigma Poisson noise error
bars. Downward pointing arrows indicate that the Poisson noise would
move these points below the contamination level as determined from the
average contaminant (Galactic stars, unresolved galaxies, image
defects) density in the outermost major and minor axis fields.
%By lining up the RGB star count profiles with luminosity profiles
%extracted from the SDSS image, we find that we reach about 31 $i_{\rm
%  AB}$-mag arcsec$^{-2}$ equivalent surface brightness.

Most conspicuous in the comparison of the different midplane profiles
is that they all show a break in their exponential slope at about
$-420$\arcsec\ ($-9.0$ kpc), independent of stellar population. {\it
  Whatever causes this truncation, it affects stars of very different
  ages in the same way.} The projected scale length in the midplane
changes by at least a factor 4 for RGB and AGB stars, and even more so
for the young stars as they have a much larger inner scale
length. This flatter distribution for young populations could be
caused by star formation mainly occurring in a ring or in wide spiral
arms. The radial distribution is not symmetric in both directions ---
especially for young stars, supporting the aforementioned spiral arm
interpretation --- and our South-West field is not at a large enough
radius to detect the break on that side.
%from 135 to 35 arcsec.
%The projected scale length of AGB and RGB stars changes at
%the break from ***\arcsec\ (***kpc) to about ***arcsec (***kpc). The
%intrinsic scale size beyond the break could be even shorter as the
%edge-on projection will blur the break. 

The MS and RGB populations have steep midplane surface density profiles beyond
the break, dropping by at least a factor of 600 in 5 kpc.
For the YMS and AGB profiles we cannot access the exact behavior
across the break because of the much lower counts so that the
profiles get lost in the gap between two ACS fields at $-500$\arcsec\ to
$-600$\arcsec. However, the upper limits in the outer fields show that
a significant break must occur for these populations as well.
%These stars are the
%brightest in our fields and we are essentially complete in the outer
%fields, therefore they are just not present at the 0.2 star per arcmin
%square level.

The RGB star count profiles agree well with luminosity profiles
extracted from the SDSS image, and by lining them up, we find that we
reach about 31 $i_{\rm AB}$-mag arcsec$^{-2}$ equivalent surface
brightness.  Our measurements also agree well with previous integrated
light measurements by \citet{vdKSea81a} and \citet{Fry99}. These
authors find a truncation radius of about 570\arcsec, larger than our
420\arcsec\ break radius. However, their truncation measurements are
the average of the shorter North-East break radius observed here and
the larger South-West break radius, and the edge-on projection of the
truncation causes deviations from an exponential profile at radii
smaller than the truncation radius.

At about $-640$\arcsec\ the MS and RGB star
count profiles seem to change scale length again to a shallower slope. This
occurs close to our detection limit and the shape and slope of this
outer profile is poorly constrained due to the uncertainty in the
contamination level. This might be a diffuse halo, akin to our detection
on the minor axis \citep{Seth07}, as this component shows
only marginal concentration toward the midplane. 

Figure~\ref{NGC4244majpop} also highlights the behavior of the
break as a function of height above the midplane.
Inside the break, the profiles at each of the vertical offset
positions have very similar radial distributions for a given
population. The density differences between the different vertical
positions are larger for the younger populations due to their shorter
scale height \citep{Seth05II}.  The break occurs at roughly the same
radius in the 50\arcsec\ offset profiles as in the midplane
profiles. Beyond -420\arcsec\ the slope of the profile gets shallower
because all profiles beyond the break, independent of the height above
the disk, converge to the same star count level at about 640\arcsec\
from the center. At 80\arcsec\ above and below the disk there is no
longer a break visible as the profile slope beyond the break has
become equal to the inner slope.

\section{Discussion and Conclusions}
\label{concl}

Our HST/ACS star count data for NGC\,4244 allow us to derive the
following observations:\\
%\begin{itemize}
%\item 
i) The North-East major axis radial star count profiles show a
  sharp break with two exponential sections. The stellar surface density
  drops by at least a factor of 600 before reaching a diffuse halo or
  the background contamination level. 
%, or at least 6 magnitudes in
%  equivalent surface brightness.
%\item 
  \\
  ii) The surface density profiles on the major axis for the young,
  intermediate age, and old populations all show a break at the same
  radius.
%\item 
\\
iii) The break occurs at the same radius independent of height above
  the midplane. The inner scale length is relatively constant, but the
  outer slope becomes gradually shallower with increasing disk
  height. This means that in face-on projections the break will be
  less pronounced.
%\item 
%\\
%iv) We find clear evidence for an extended component along the minor
%  axis of NGC\,4244 with a larger scale size than the inner vertical
%  exponential component. We find suggestive evidence that a similar
%  component is present on the major axis at similar surface
%  densities. Beyond the break this component shows no concentration
%  toward the midplane. 
%The exact shape of this component cannot be
%  determined due to the uncertainty in contamination and low number
%  statistics. 
%\item 
%\\
%v) The young population is more concentrated toward the midplane
%  than the intermediate and old populations, independent of radius.
%\end{itemize}

The observation that all stellar populations, regardless of age,
have similar breaks in the midplane profiles puts the strongest
constraint on models of break formation, assuming that the
NGC\,4244 break is representative of other breaks.  Most
notably, any model where breaks are caused by a sudden change in
star formation rate will now have to explain why the radius at which
this occurs is stable over a very
long time. The model will also have to explain why the breaks in
the different populations have similar shapes, whereas one would expect
that a very old break would be smoothed by the dynamical heating that we
know occurs in older populations.
%, also for NGC\,4244 \citep{Seth05II}.

For instance, a model like that of \citet{vdK87}, in which the break
results from the specific angular momentum distribution of a
collapsing gaseous sphere that limits the maximum angular momentum
available to redistribute gas in the outer disk, has several
problems. Over time, more and more gas will decouple from the Hubble
flow with different distributions of specific angular momentum, and it
is hard to conceive that the cutoff radius would never
change. Furthermore, the formation of bars and spiral arms
redistributes angular momentum from the central regions to the outer
disks so that the current specific angular momentum distribution no
longer necessarily reflects the initial distribution
\citep[e.g.,][]{Deb06}. Finally, as the \hi\ gas distribution of many
galaxies is significantly more extended than their stellar
distribution, it would seem that higher angular momentum gas is often
already present \citep[although this is currently not true for
NGC\,4244; ][]{Oll96}.

Models that rely on some star formation threshold will require fine
tuning to keep the threshold at the same location with respect to the
broad stellar distribution. Models that rely on the properties of the
gas alone (e.g., \citealt{Sch04} for gas density; \citealt{Ken89} for
gas stability) will need to explain why any accreted gas redistributes
itself in such a way as to keep the threshold in place for 10 Gyr.
% (ongoing gas
%accretion is necessary because most spiral galaxies would otherwise
%exhaust their gas reservoir within a few Gyr). 
It is certainly possible that the gas responds to the stellar mass
already present, as gas flaring and warping seems to indicate
\citep{vdK07}. In NGC\,4244, the \hi\ rotation curve drops as the gas
density begins to decline, flare, and warp all at about 550\arcsec\ 
\citep{Oll96}, somewhat beyond the break radius.
%, but a
%simple threshold based on gas stability criteria
%\citep[e.g.,][]{Ken89}, or gas surface density \citep[e.g.,][]{Sch04} will need
%further tuning as the threshold location in these models will almost
%certainly change with time as the galaxy accretes more mass over
%time. 
A model that depends on both gas and stellar mass distribution
\citep[e.g., ][]{DopRyd94} might provide a solution, but still would
have a hard time explaining the sharpness of the break in the very old
RGB population.  Furthermore, all these models may have problems
explaining why not all galaxies show breaks (e.g., NGC\,300 has an
exponential profile over 10 scale lengths; \citealt{BlH05}).

Secular evolution driven by bars and spiral arms can also lead to
breaks by redistributing angular momentum and hence stellar mass
\citep[e.g., ][]{Deb06}. This dynamical redistribution of mass will
be the same for all age populations. 
%and as long as star formation is
%reasonable proportional to the local (stellar) surface density (for
%which there is some evidence; \citealt{BelldeJ00}), the break and
%profile will be similar for all populations. 
However, if we want to create a break by secular evolution of stars
alone, the break must be created on a time scale of less than 100
Myr. When we account for the fact that the gas will also be involved
in the secular evolution, we can allow longer time scales, but we then have a
similar problem as with the threshold models in that star formation
must be proportional to stellar mass density. Furthermore, the secular
evolution model requires a significant bar, yet we find no evidence for a
strong bar in NGC\,4244. Our Spitzer 3.6 and 4.5 micron images
(Holwerda et al., in prep.) show an elongated shape with some twist in
the isophotes out to 50\arcsec\ (1/3 of a disk scale length), but we
see no vertical thickening. The \hi\ velocity field shows some
irregularities between 200\arcsec\ and 300\arcsec\ \citep{Oll96} that
could point to a larger bar, but it is not obviously associated with
any features in the optical and IR images. Finally, the factor 600
decrease in surface density at the break (or $\sim$100 compared to
the extended inner exponential disk) makes it unlikely that the break is
just a spiral arm seen edge-on. This amplitude of arm-interarm contrasts
has never been observed in face-on galaxies.
% Still, we
%could be seeing a bar on its long-end and it would be hard to detect.

Tidal stripping by galaxy interactions might be another route for
creating galaxy breaks (and anti-breaks). This mechanism is only
effective after multiple interactions, as in a galaxy cluster
environment \citep{Gne03}. This is unlikely to be the case here as
NGC\,4244 is the second brightest member of a very loose group with
mainly very late-type irregular members and NGC\,4244 has no detected \hi\
companions within 40 kpc \citep{Dah05}.

Disk heating and star stripping by pure dark matter sub-halos with a
$\Lambda$CDM power spectrum is the final model for disk breaks we
consider here (Kazantzidis et al., in prep.). This model has
theoretical similarities to the galaxy/cluster interaction model
described above, except that the mass power spectrum of interactions is very
different. In this model disk stars are heated by the constant
bombardment of dark matter sub-halos whizzing through the disk.
%stripping loosely bound stars on the outskirts of the galaxy. 
The
original stars in a thin disk will spread into a thicker disk and a
diffuse halo develops. At the disk outskirts, where disk
self-gravity is low, stars are stripped and a truncation develops. 
%The
%truncation will therefore most likely be at a similar radius as
%calculated above for interactions and warping \citep{SahaJog06}. 
If the stripping occurs quickly, independent of stellar age and
scale height, this model might well explain our observations.

The stripped stars would form a diffuse, very flattened halo around the
galaxy with stream-like substructure due to the interactions with the dark
matter sub-halos. This is not unlike the structure we see
around NGC\,4244, with extended components along the major and minor
axis of similar surface densities, which, if present around the whole
galaxy, would form a very flattened extension around the
disk. Unfortunately, the number of stars detected in the halo
of NGC\,4244 is too low to measure substructure.  Any stellar halo
formed through sub-halo stripping should not be confused with the
stellar halos that form from disrupted satellites in hierarchical
galaxy formation. The diffuse halo we detect here is more massive than
predicted by these kinds of models for galaxies with masses similar to
that of NGC\,4244 \citep{Pur07}.

It is conceivable that several of the above models are at
play simultaneously. To have a sharp, deep factor of 600 break in the
old RGB population one likely needs one of the dynamical methods,
in which the truncation radius is probably set where the stellar disk
self-gravity no longer dominates the potential and where warps are
predicted to start \citep{SahaJog06}. Once one of the dynamical
methods has created a truncation, later infalling gas will respond to
the overall potential in such a way as to set up a star formation
threshold at that location. Alternatively, the dynamics of a spiral
density wave may interact with a star formation threshold to set a
similar feature in the radial profile for young and old
stars. Notwithstanding this caveat, dynamical break explanations
are favored as they more naturally explain why profile shapes
should be similar for both young and old stars.  Analysis of similar
GHOSTS data on a large number of galaxies promises to strengthen the
constraints on these models.


%% Included in this acknowledgments section are examples of the
%% AASTeX hypertext markup commands. Use \url without the optional [HREF]
%% argument when you want to print the url directly in the text. Otherwise,
%% use either \url or \anchor, with the HREF as the first argument and the
%% text to be printed in the second.

\acknowledgments

Support for Proposal numbers 9765 and 10523 was provided by NASA
through a grant from the Space Telescope Science Institute, which is
operated by the Association of Universities for Research in Astronomy,
Incorporated, under NASA contract NAS5-26555.


%% in parentheses, after the keyword, but they will not be verified.

{\it Facilities:} \facility{HST (ACS)}, \facility{Sloan}.




%\begin{thebibliography}{}
\bibliographystyle{apj}
%\bibliography{apj-jour,myrefs}

%\end{thebibliography}

\end{document}